\documentclass[conference]{IEEEtran}
\IEEEoverridecommandlockouts
\usepackage{cite}
\usepackage{amsmath,amssymb,amsfonts}
\usepackage{algorithmic}
\usepackage{graphicx}
\usepackage{booktabs}
\usepackage{multirow}
\usepackage{textcomp}
\usepackage{siunitx}
\usepackage{xcolor}
\usepackage{soul, color}
\usepackage{algorithm}
\usepackage{booktabs}
\usepackage{fancyhdr}
\def\BibTeX{{\rm B\kern-.05em{\sc i\kern-.025em b}\kern-.08em
    T\kern-.1667em\lower.7ex\hbox{E}\kern-.125emX}}
\begin{document} 

\title{Memory-Based Set Point Modulation for Improved Transient Response of Distributed Energy Resources \\
\thanks{
© 20XX IEEE.  Personal use of this material is permitted.  Permission from IEEE must be obtained for all other uses, in any current or future media, including reprinting/republishing this material for advertising or promotional purposes, creating new collective works, for resale or redistribution to servers or lists, or reuse of any copyrighted component of this work in other works.

This work is supported in part by the National Science Foundation (NSF) under award ECCS-1953213, in part by the U.S. Department of Energy's Office of Energy Efficiency and Renewable Energy (EERE) under the Solar Energy Technologies Office Award Number 38637 (UNIFI Consortium led by NREL), in part by Manitoba Hydro International, and in part by the Commonwealth Cyber Initiative, an investment in 
the advancement of cyber R\&D, innovation, and workforce development. For more information about CCI, visit www.cyberinitiative.org. The views expressed herein do not necessarily represent the views of the U.S. Department of Energy or the United States Government.
}
}

\author{
\IEEEauthorblockN{
Milad Beikbabaei, Brady Alexander, Ashwin Venkataramanan,  and Ali Mehrizi-Sani 
}
\IEEEauthorblockA{The Bradley Department of Electrical and Computer Engineering\\
Virginia Polytechnic Institute and State University, Blacksburg, VA 24061\\
Emails:\{miladb, bradyba19,vashwin, mehrizi\}@vt.edu}
}

\maketitle
\begin{abstract}
As the composition of the power grid evolves to integrate more renewable generation, its reliance on distributed energy resources (DER) is increasing. Existing DERs are often controlled with proportional integral (PI) controllers that, if not properly tuned or if system parameters change, exhibit sluggish performance or large overshoot. The use of set point automatic adjustment with correction-enabled (SPAACE) with a linear predictor improves the transient response of these DERs without the need to access the PI controller parameters.  
The limitation of the existing SPAACE method is the high sampling rate needed for improved performance, which is not always practical. 
This paper proposes the addition of a memory term to the SPAACE with a linear predictor. This memory term is the integral of the errors of previous samples, which adds another layer to the prediction to improve the response at lower sampling rates and further reduces the overshoot and settling time compared to the existing SPAACE method. Time-domain simulation studies are performed in PSCAD/EMTDC to show the effectiveness of the proposed controller.
\end{abstract}


\section{Introduction}

The power system is transitioning to renewable generations, and more inverter-based resources (IBR) and distributed energy resources (DER) are connecting to the grid. Integration of IBRs and DERs reduces the mechanical inertia, decreases the overall grid strength, and adds faster transients to the grid~\cite{low_inertia,inertia_challange,power_system_stability_etal, DER_inertia,milad_FDI_LSTM}.
As a result of these changes, the existing DERs may show poor performance. Previous work has studied the impact of the aforementioned changes on the performance and stability of DERs~\cite{DER_stability,low_inertia}.  
As a result, there is a need for a method that improves the performance of existing DERs~\cite{DRLS, blackbox_systemidentificaitonPV, advance_control,DER_control}. 
However, retuning existing DERs to improve their performance is not always possible to the proprietary rules from DER manufacturers and vendor warranties. As a result, retuning internal gains or redesigning internal controls in these DERs is not preferred. A black-box control scheme that operates outside the periphery of DERs without modifying internal control settings is desired.

Set point automatic adjustment with correction-enabled (SPAACE)~\cite{SPAACE_theory} is a technique in literature that enables black-box control in DERs. SPAACE shapes the reference signal or set point issued to a DER to improve its transient performance. SPAACE has proven effective in a wide range of applications. SPAACE has also been previously tested in laboratory experimental setups~\cite{SPAACE_practical}.
Predictive set point modulation is developed for energy storage systems to reduce the overshoot and the settling time in~\cite{SPM_ESS}.
References~\cite{coordinated_SPM,choudhuryBiplav2023} add a communication layer to SPAACE to show improved transients in low inertia microgrids.
Reference~\cite{CSPAACE_5G} implements SPAACE with communication using an RTDS testbed co-simulated with a {5G} testbed, where SPAACE with communication decreases the settling time during the faults compared to SPAACE without communication.
A linear predictor is often utilized to predict the magnitude of response a few times steps ahead~\cite{SPAACE_theory}, which is further utilized in shaping the reference signal. 
The performance of the SPAACE can be improved by utilizing a faster observer or sensor. However, this places additional infrastructure costs or may not even be feasible.
A predictive set point modulation control method is developed for supercapacitors in a DC grid~\cite{Supercap_SPM}.
SPAACE is combined for controller current to reduce the current overshoot for parallel charging system~\cite{SPAACE_4_parallel_charging}.
Reference~\cite{DRLS} uses an optimization-based approach, where a reinforcement learning (RL)--based algorithm determines optimal set point discrete set points based on local measurements without utilizing a predictor. 
%

SPAACE with a linear predictor~\cite{SPAACE_theory} works considerably well for slow transients, and the sampling rate of the observer needs to be higher with faster transients to improve the prediction accuracy. At slow sampling rates, the performance of the strategy reduces or even results in unstable behavior. 
Furthermore, SPAACE performance depends on the grid short circuit ratio (SCR), where a strong grid usually has an SCR of greater than 5~\cite{hadavi2022}.
This paper proposes the addition of a memory term to the linear predictor in SPAACE; the memory term is calculated using the integral of the error of the previous samples.
The use of the memory term increases the performance of SPAACE since the observer accuracy increases from using a window of past data rather than only using the current input. The proposed method, SPAACE with memory (SPAACE-M), improves the performance of SPAACE with a linear predictor for slower sampling time, with the following salient features:
    \begin{itemize}
    \item Works with a lower sampling rate, as low as 3~ms.
    \item Improves SPAACE performance for fast transient responses, with respect to the overshoot, undershoot, and settling time, for a fixed sampling rate.
    \item Works for both a weak and strong grid.
    \end{itemize}
The rest of the paper is structured as follows. Section~II discusses the formulation, benefits, and limitations of the black-box set point control method, SPAACE with a linear predictor~\cite{SPAACE_theory}. Section~III provides an overview of the proposed SPAACE-M controller. In Section~IV, the proposed method is tested utilizing electromagnetic transient (EMT) simulations in PSCAD/EMTDC software. Finally, conclusions are presented in Section~V.


\section{Set point automatic adjustment with correction-enabled (SPAACE)}

\begin{figure}[!t]
\centering
\includegraphics[width=1\columnwidth]{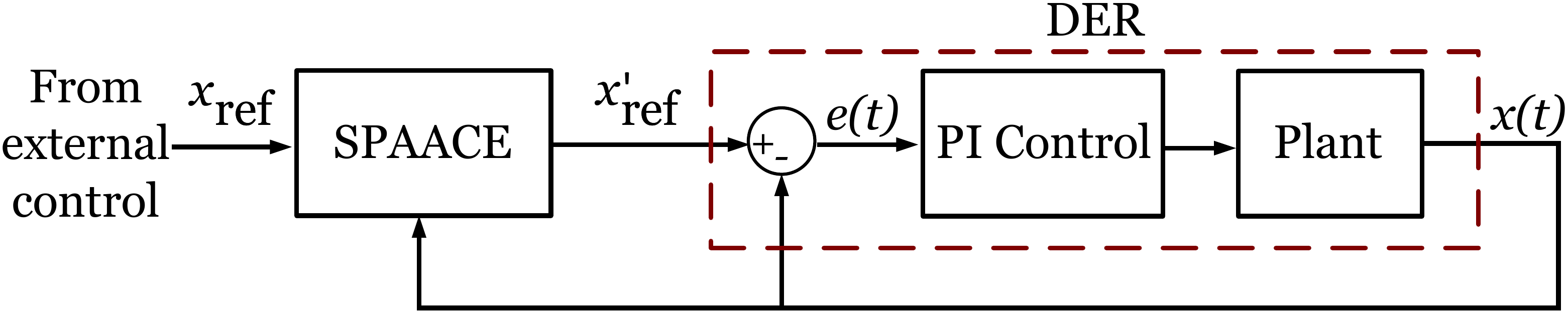}
\caption{High-level schematic of SPAACE.}
\label{fig:SPAACE_schematic}
\end{figure}

This section discusses the existing control method, SPAACE with a linear prediction. 
As mentioned in Section~I, with SPAACE, the reference signal issued to the control process is manipulated temporarily to improve its transient performance. Fig.~\ref{fig:SPAACE_schematic} shows the block diagram representation of SPAACE connecting to a DER. SPAACE receives the set point from an external higher-level controller and issues a modified set point to the controlled plant. Since the strategy only requires the issued set point and the feedback variable to achieve its control objective, it can be utilized to enable black-box control in grid-following (GFL) DERs.

In order to modify the issued set point, the strategy utilizes a linear predictor to predict the tracking error $e$ a few steps ahead:
\begin{equation}
e(t_k) = x_\text{ref}(t_k) - x{(t_k)},
\label{eq:error}
\end{equation}
where $x_\text{ref}$ is the input reference and $x{(t)}$ is the measured output at $t = t_k$.
The prediction time step $T_\text{sample}$ can be any integer multiple of the sampling rate $T_\text{sample}$ as 
\begin{equation}
T_\text{pred} = n \: T_\text{sample},
\label{eq:1}
\end{equation}
where $T_\text{sample}$ is the observer sampling time and $n$ is adjusted based on the frequency of the unmanipulated transient response.  
Furthermore, the predictor operates based on the rate of change $r(t_k)$ of the controlled variable $x$ over a predicted interval $t_\text{pred}$, where $r(t_k)$ is

\begin{equation}
r(t_k) = \frac{x{(t_k)} - x{(t_k - T_\text{pred} )} }{T_\text{pred}}.
\label{eq:4}
\end{equation}

The predicted input $x{(t_k+T_\text{pred})}$ is

\begin{equation}
x{(t_k+T_\text{pred})} = x{(t_k)} + r(t_k)\:T_\text{pred}.
\label{eq:5}
\end{equation}

Utilizing $x{(t_k+T_\text{pred})}$ the predicted error $e_\text{pred}(t_k)$ is 
\begin{equation}
e_\text{pred}(t_k) = x_\text{ref}(t_k) - x{(t_k+T_\text{pred})}.
\label{eq:errorpredicted}
\end{equation}

Finally, the strategy issues temporary modifications to the set point $x'_\text{ref}$ 

\begin{equation}
x'_\text{ref}(t_k)= 
\begin{cases}
\begin{aligned}

    & x_\text{ref}(t_k) + m\:e_\text{pred}(t_k),     &|e(t)|>0.05\\
    &x_\text{ref}(t_k),               &\text{otherwise},

\label{final_equation_SPAACE}
\end{aligned}
\end{cases}
\end{equation}
where $m$ is simply a scaling factor for the predicted error term, and the method starts issuing set point changes if the error is higher than a small value, say 0.05~pu in this case. The parameters $m$ and $t_\text{pred}$ are usually tuned similar to the parameters of a PI controller and can be identified through trial-and-error. The pseudocode of SPAACE with a linear predictor is shown in Fig.~\ref{pseudocode-SPAACE}.

\begin{figure}
 \begin{algorithm}[H]
\caption{SPAACE pseudocode.}
 \begin{algorithmic}[1]
  \STATE $T_\text{pred} = n \: T_\text{sample}$ 
  \STATE $e(t_k) = x_\text{ref}(t_k) - x{(t_k)}$
  \STATE $r(t_k) = \frac{x{(t_k)} - x{(t_k - T_\text{pred} )} }{T_\text{pred}}$
  \STATE $x{(t_k+T_\text{pred})} = x{(t_k)} + r(t_k)\:T_\text{pred}$
  \STATE $e_\text{pred}(t_k) = x_\text{ref} - x{(t_k+T_\text{pred})}$
   \IF{$|e(t_k)|>\epsilon$}
     \STATE $x'_\text{ref}(t_k) = x_\text{ref}(t_k) + m\:e_\text{pred}(t_k) $
  \ELSE
    \STATE  $x'_\text{ref}(t_k) = x_\text{ref}(t_k)$
  \ENDIF
 \end{algorithmic}
 \end{algorithm}
 \caption{Pseudocode of SPAACE algorithm. }
 \label{pseudocode-SPAACE}
\end{figure}


\begin{figure}[!t]
\centering
\includegraphics[width=1\columnwidth]{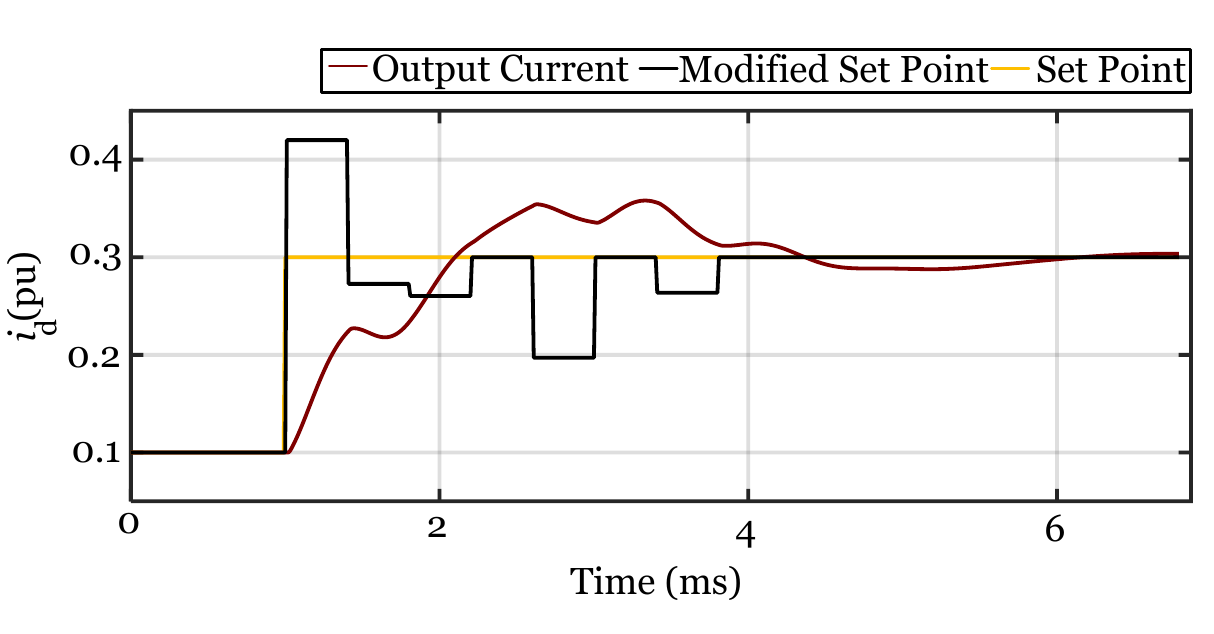}
\caption{Set point modulation with SPAACE for a step change in $i_d$ from 0.1~pu to 0.3~pu at $t=1~\rm{ms}$.}
\label{concept_SPAACE}
\end{figure}

Fig.~\ref{concept_SPAACE} shows an example case of a generic current measurement $i_d$ to show how SPAACE modulates a set point to improve the transient response. At $t=1~\rm{ms}$, a step change increases the $i_d$ set point from 0.1 to 0.3~pu. Instead of stopping at the new set point, SPAACE increases its set point above 0.41~pu to pull the output current measurement up quicker to its final value. Once the output current overshoots the final value, SPAACE modulates the set point under 0.3~pu to bring the output current down to the desired final value.



\section{ SPAACE with Memory (SPAACE-M)}
This section discusses the proposed SPAACE with memory (SPAACE-M) method. SPAACE-M builds on the SPAACE with a linear predictor method~\cite{SPAACE_theory} and adds the integral of the error measured from previous samples. The integral is implemented by averaging a window of the $j$ past tracking error $e$ at each simulation time step. The integral formula is
\begin{equation}
e_\text{past}(t_k) = \frac{\sum^{j}_{i=0}{e(t_k - i\:T_{\text{sample}})} } {j}.
\label{eq:errorpredicted}
\end{equation}

Utilizing the accumulative error and the next time step prediction output, the strategy in~\eqref{final_equation_SPAACE} can be reconfigured as follows, 

\begin{equation}
x'_\text{ref}(t_k)= 
\begin{cases}
\begin{aligned}

    & x_\text{ref} + m_{1}\:e_\text{pred}(t_k) + m_{2}\:e_\text{past}(t_k),     &|e(t)|>0.05\\
    &x_\text{ref}(t_k),               &\text{otherwise},
\label{eq:4}
\end{aligned}
\end{cases}
\end{equation}
where $e_\text{pred}$ is the future predicted error, while $e_\text{past}$ is the average of the window of $j$ previous errors, in this paper $j=2$. 
Furthermore, $m_{1}$ and $m_{2}$ are the weights associated with each error. Using the implementation of the current controller of a DER as an example, if the tracking error is higher than a threshold $\epsilon$, the strategy starts adjusting the issued reference $i^*$ to a new value $i_d^{'*}$ based on~\eqref{eq:4} at each time step. The action is repeated till the tracking error is reduced below 0.05 as shown in (8) above. The pseudocode implementation of the SPAACE-M algorithm is shown in Fig.~\ref{pseudocode_SPAACEM}.

\begin{figure}
 \begin{algorithm}[H]
\caption{SPAACE-M pseudocode.}
 \begin{algorithmic}[1]
   \STATE $T_\text{pred} = n \: T_\text{sample}$ 
  \STATE $e(t_k) = x_\text{ref}(t_k) - x{(t_k)}$
  \STATE $r(t_k) = \frac{x{(t_k)} - x{(t_k - T_\text{pred} )} }{T_\text{pred}}$
  \STATE $x{(t_k+T_\text{pred})} = x{(t_k)} + r(t_k)\:T_\text{pred}$
  \STATE $e_\text{pred}(t_k) = x_\text{ref} - x{(t_k+T_\text{pred})}$
   \STATE $e_\text{past}(t_k) = \frac{\sum^{j}_{i=0}{e(t_k - i\:T_{\text{sample}})} } {j} $
   \IF{$|e(t_k)|>\epsilon$}
     \STATE $x'_\text{ref}(t_k) = x_\text{ref}(t_k) + m_{1}\:e_\text{pred}(t_k) + m_{2}\:e_\text{past}(t_k)$
  \ELSE
    \STATE  $x'_\text{ref}(t_k) = x_\text{ref}(t_k)$
  \ENDIF
 \end{algorithmic}
 \end{algorithm}
 \caption{Pseudocode of SPAACE-M algorithm. }
 \label{pseudocode_SPAACEM}
\end{figure}


Similar to SPAACE with linear prediction, $m_{1}$ and $m_{2}$ must be chosen through trial-and-error. 
$m_{1}$ is the gain of the next time step prediction, and $m_{2}$ is the gain for the average of the previous errors.
When $m_{1}$ is large, the method is capable of prioritizing short-term disturbances, leading to a stronger response to short-term disturbances. Furthermore, a larger $m_{2}$ results in prioritizing the previous error, increasing the accuracy of the method.

\section{Performance Evaluation}

\begin{figure}[!t]
\centering
\includegraphics[width=0.9\columnwidth]{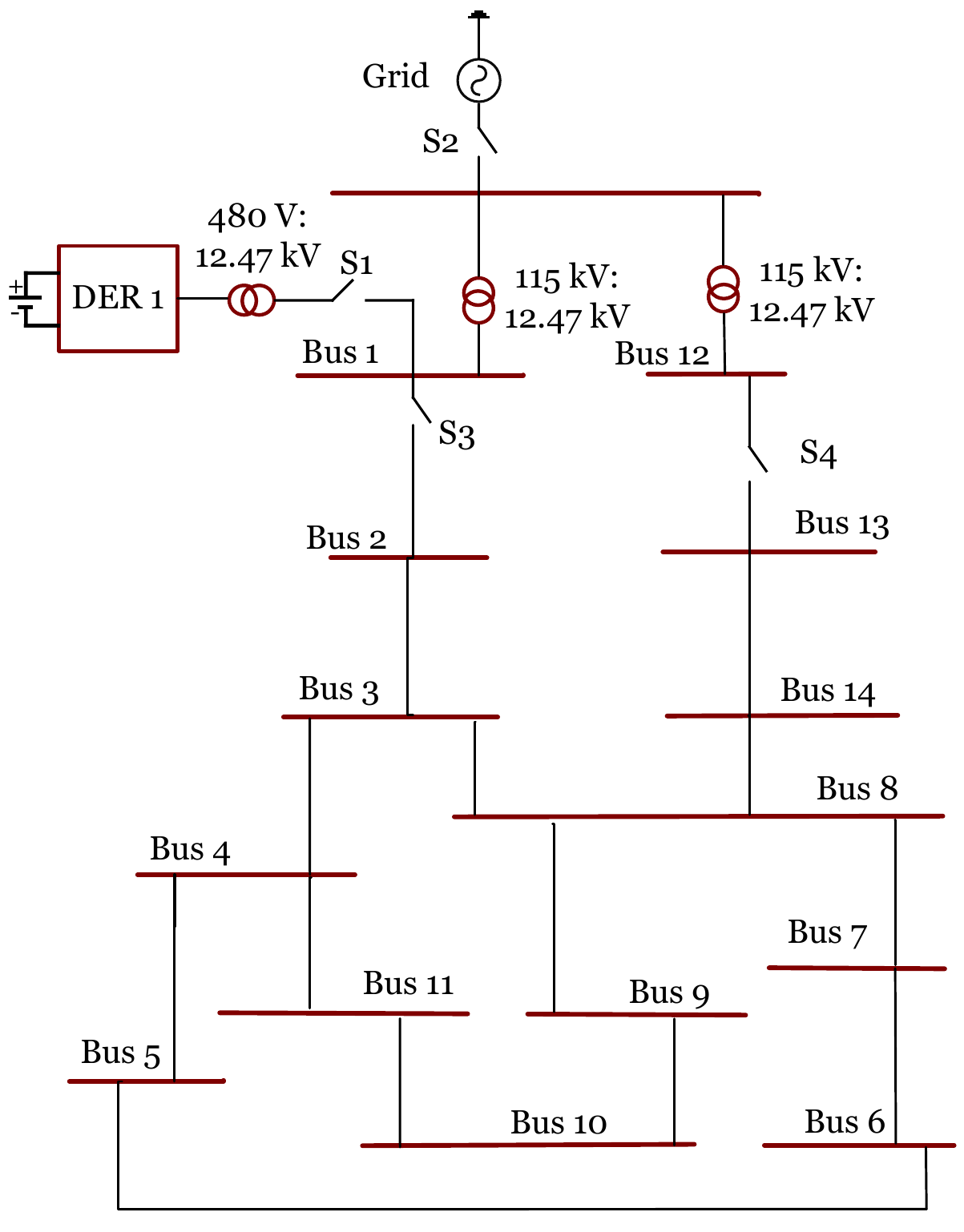}
\caption{Modified CIGRE 14-bus north American benchmark test system.}
\label{fig:CIGRE-14}
\end{figure}

\begin{figure}[!t]
\centering
\includegraphics[width=1\columnwidth]{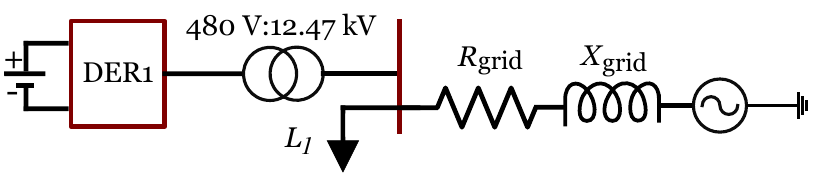}
\caption{A DER connected to a grid through a transformer.}
\label{single_bus_system}
\end{figure}

The performance of SPAACE and SPAACE-M are compared under the current reference step change and a three-phase short-circuit fault with different sampling rates using the modified 14-bus CIGRE North American benchmark shown in Fig.~\ref{fig:CIGRE-14}~\cite{CIGRE-14}.
In the modified CIGRE test system, the loads are distributed across all phases to remove any imbalance. A 6~MVA DER operates in the GFL mode at bus~1. 
Moreover,  SPAACE and SPAACE-M are tested under different grid strengths using a single bus system shown in Fig.~\ref{single_bus_system}. Grid strength is an evaluation of how resilient the grid is to disturbances and is characterized by short circuit ratio (SCR). A weak grid is considered to have an SCR lower than 3, and a stiff grid is considered to have  an SCR greater than 5~\cite{hadavi2022}. 

For all cases, the SPAACE and SPAACE-M logic are implemented in Python using the PSCAD co-simulation block~\cite{CoSimulationPSCAD_milad}. 
The DER operates only in GFL mode, receiving a $dq$ current reference $i_{d,ref}$ and $i_{q,ref}$. SPAACE and SPAACE-M take in $i_{d,ref}$ as their inputs and generate a modulated reference $i_{d,ref}^*$. Furthermore, SPAACE and SPAACE-M are compared against the base case where the proportional integral (PI) controller of the DER receives $i_{d,ref}$ directly. In all of the following cases, the future scaling factor is $m_1 = -0.3$, the memory factor is $m_2 = -1$, $\epsilon=0.05$, and $t_\text{pred} = 0.8~\rm{ms}$. These values are selected through trial and error.

SPAACE and SPAACE-M exhibit the greatest difference in performance under varying sampling times, for transient events such as faults, and when the strength of the grid varies. 
The following sections show these cases, showing the gains that result from adding a memory term to the SPAACE with a linear predictor method~\cite{SPAACE_theory}.

\subsection{Case 1: Impact of Sampling Rate}

\subsubsection{Case 1.1: Set Point Increase with High Sampling Rate}\label{Case_1_1}

\begin{figure}[!t]
\centering
\includegraphics[width=1\columnwidth]{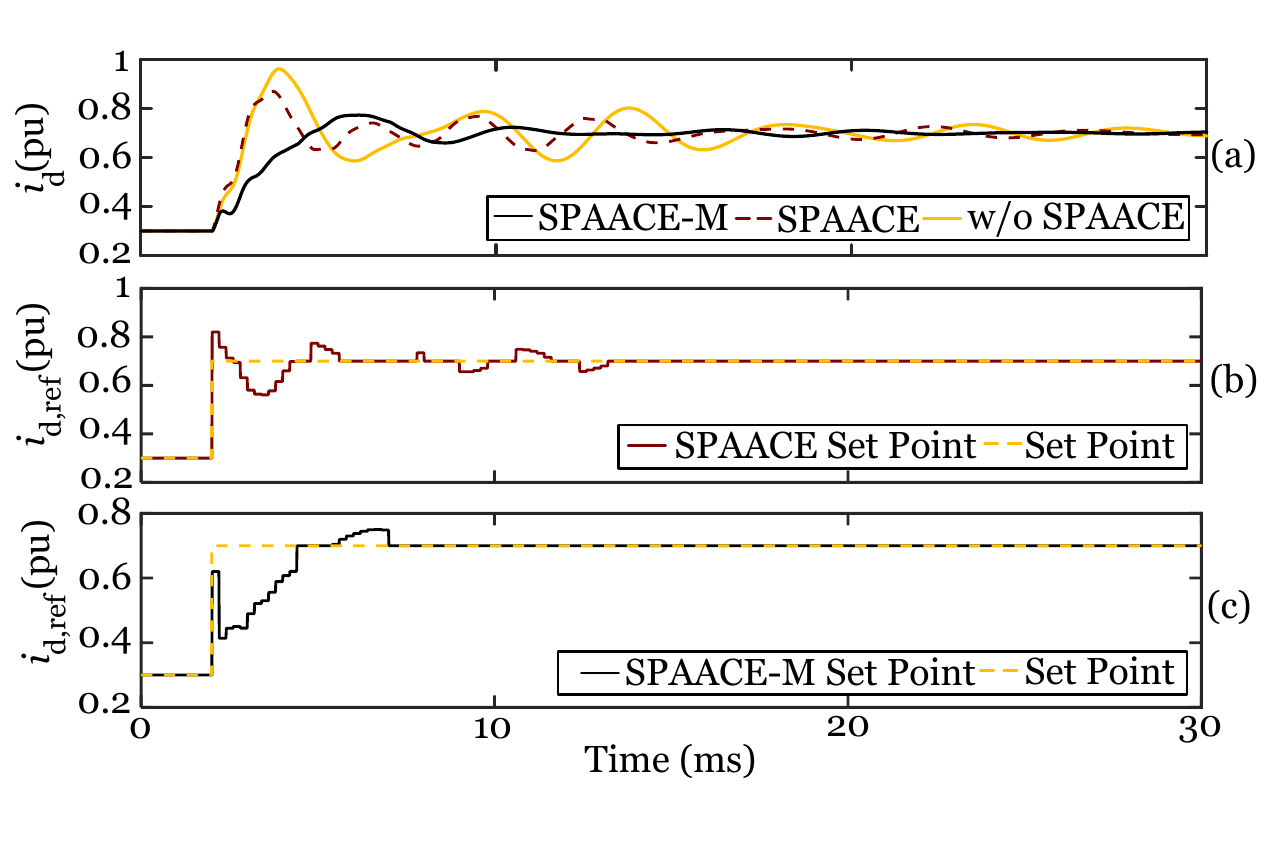}
\caption{Simulation results for Case~1.1: $i_d$ step change at $t=2~\rm{ms}$ from 0.3 to 0.7~pu with a high sampling rate.}
\label{step_change}
\end{figure}

Fig.~\ref{step_change} shows the simulation results for Case 1.1 tested on the system shown in Fig.~\ref{fig:CIGRE-14}. The DER at bus~1 is operating in the steady state before there is a step change in $i_{d,ref}$ from 0.3~pu to 0.7~pu at $t=2~\rm{ms}$. The sampling rate for SPAACE and SPAACE-M is 0.2~ms; this is considered to be a high sampling rate. Fig.~\ref{step_change}(a) shows the $i_d$ current measurements. In the base case without SPAACE, the current overshoots its final value by 37.36\% before reaching its new steady state value. SPAACE reduces the overshoot to 24.27\%, and SPAACE-M further reduces the overshoot to 10.42\%. However, SPAACE-M exhibits a slower rise time as compared to both SPAACE and the base case. Fig.~\ref{step_change}(b) shows the set point and the modulated set point given by SPAACE. 
The SPAACE modulated set point overshoots the unmodified set point before oscillating around the final value to decrease the underdamped response of the current. Fig.~\ref{step_change}(c) shows the set point and the modulated set point produced by SPAACE-M. The modulated SPAACE-M set point rises slower to 0.7~pu with only a small overshoot before settling to its final value. 

\subsubsection{Case 1.2: Set Point Increase with Low Sampling Rate}

\begin{figure}[!t]
\centering
\includegraphics[width=1\columnwidth]{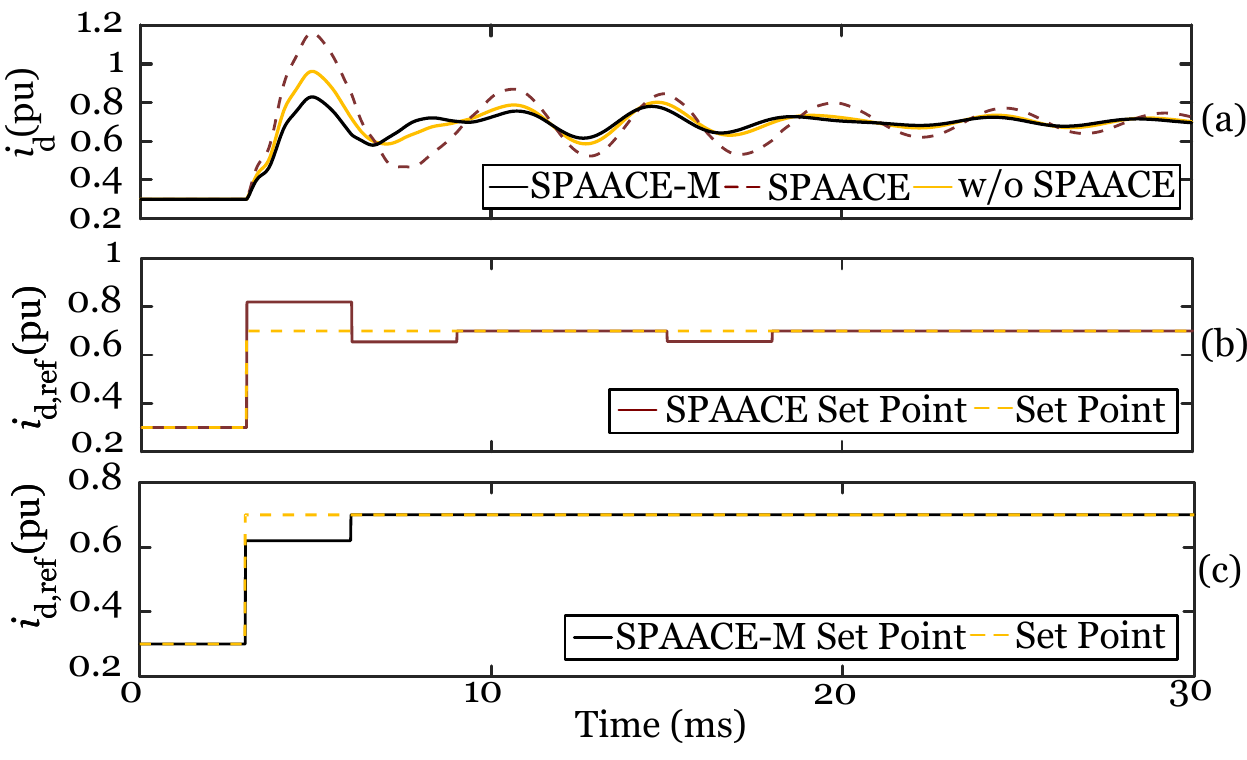}
\caption{Simulation results for Case~1.2: $i_{d,ref}$ step change at $t=2~\rm{ms}$ from 0.3 to 0.7~pu with a low sampling rate.}
\label{increasedSamplingTime}
\end{figure}

Fig.~\ref{increasedSamplingTime} shows the simulation results for Case 1.2 tested on the system shown in Fig.~\ref{fig:CIGRE-14}. The DER at bus~1 is operating in the steady state before $i_{d,ref}$ is adjusted with a step change from 0.3 to 0.7~pu $t=2~\rm{ms}$. In this case, the sampling rate for the SPAACE and SPAACE-M controllers is reduced to 3~ms, reducing the frequency that these controllers modulate the set point. Fig.~\ref{increasedSamplingTime}(a) shows the $i_d$ current measurements. In the base case, the current rises quickly to the final value with an overshoot of 37.36\%. SPAACE has a longer rise time with the increased sampling rate and an overshoot of 65.74\%. This is because the duration that the modulated set point is greater than the unmodified set point increases with the increase in sampling time, causing a greater overshoot. SPAACE-M shows a faster rise time to Case~1.1 and an overshoot of 18.45\%. Fig.~\ref{increasedSamplingTime}(b) shows the modulated set point for SPAACE. Fig.~\ref{increasedSamplingTime}(c) shows the modulated set point for SPAACE-M. SPAACE-M increases the time it takes for its set point to reach the final value, which decreases the overshoot, as compared to SPAACE and the base case. 

\subsection{Case 2: Three-Phase Line to Ground Fault}
\begin{figure}[!t]
\centering
\includegraphics[width=1\columnwidth]{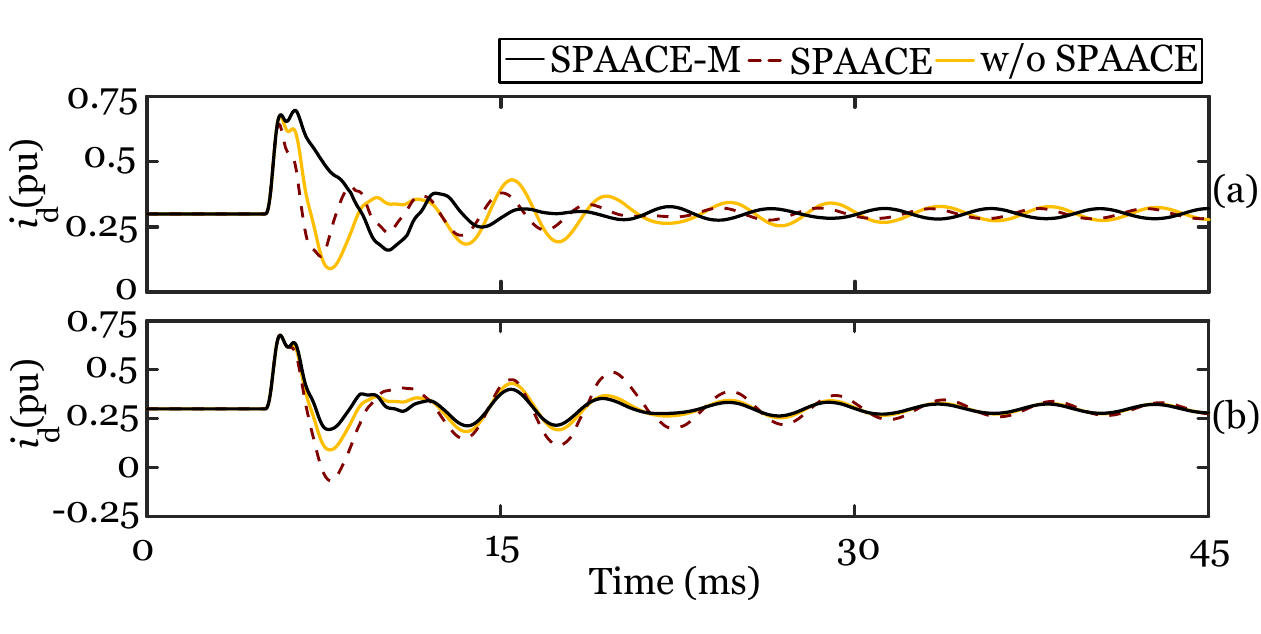}
\caption{Simulation results for Case~2: Three-phase bolted fault at $t=5~\rm{ms}$ with: (a) 0.2~ms sampling time; (b) 3~ms sampling time.}
\label{3PhaseFault}
\end{figure}

Fig.~\ref{3PhaseFault} shows the simulation results for Case 2 tested on the system shown in Fig.~\ref{fig:CIGRE-14}. The DER at bus~1 is operating in the steady state before a three-phase bolted fault occurs at bus~1. Fig.~\ref{3PhaseFault}(a) shows the fault performance for the base case, SPAACE, and SPAACE-M when the sampling rate is 0.2~ms. The SPAACE controller restores the fault current near its nominal range quicker than both the base case and SPAACE-M. SPAACE-M shows a longer duration of fault current at the maximum but decreases the undershoot as compared to both SPAACE and the base case. Fig.~\ref{3PhaseFault}(b) shows the fault performance for the base case, SPAACE, and SPAACE-M when the sampling rate is 3~ms. All 3 cases have similar overshoot, but SPAACE shows the greatest undershoot. SPAACE-M reduces the undershoot, restoring $i_d$ faster than the other two cases.

\begin{figure}[!t]
\centering
\includegraphics[width=1\columnwidth]{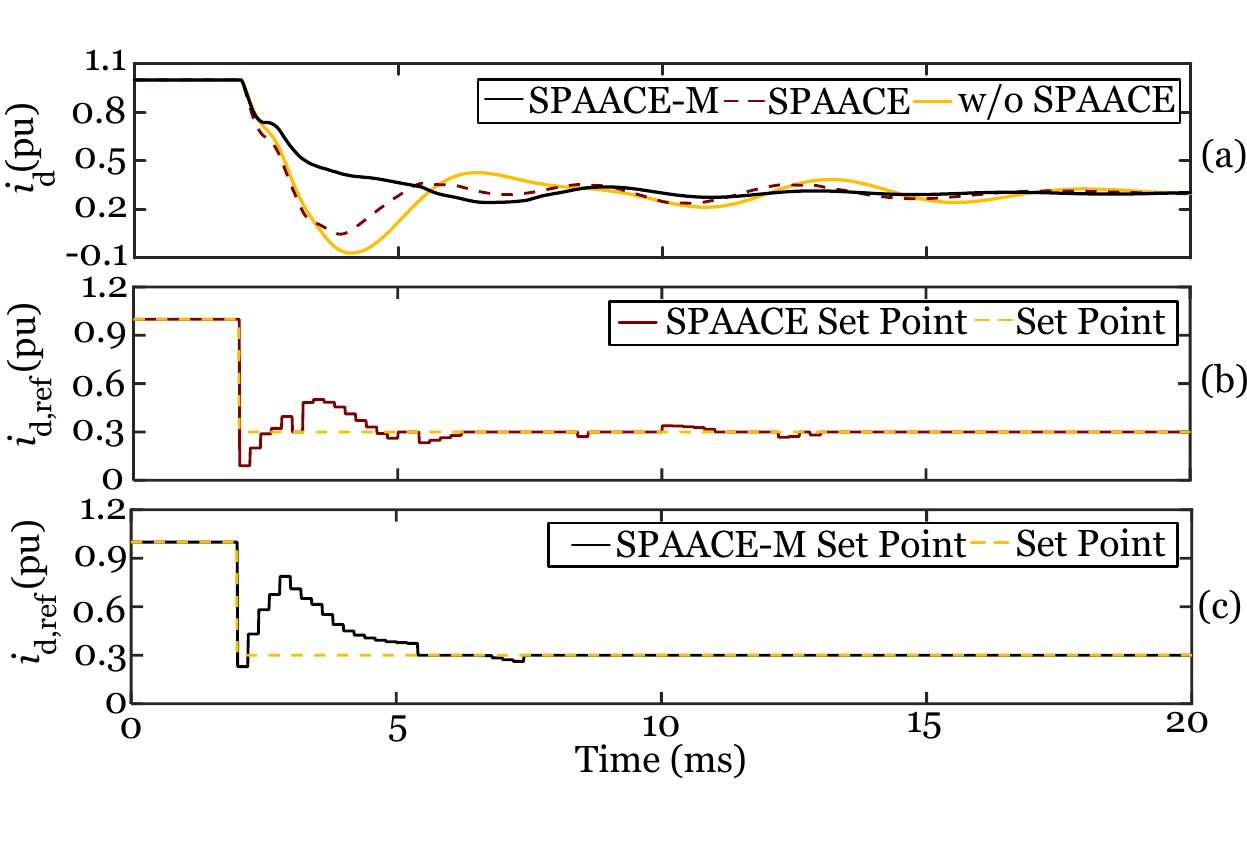}
\caption{Simulation results for Case~3.1: $i_d$ step change at $t=2~\rm{ms}$ from 1 to 0.3~pu with SCR = 5.}
\label{scr5}
\end{figure}

\begin{figure}[!t]
\centering
\includegraphics[width=1\columnwidth]{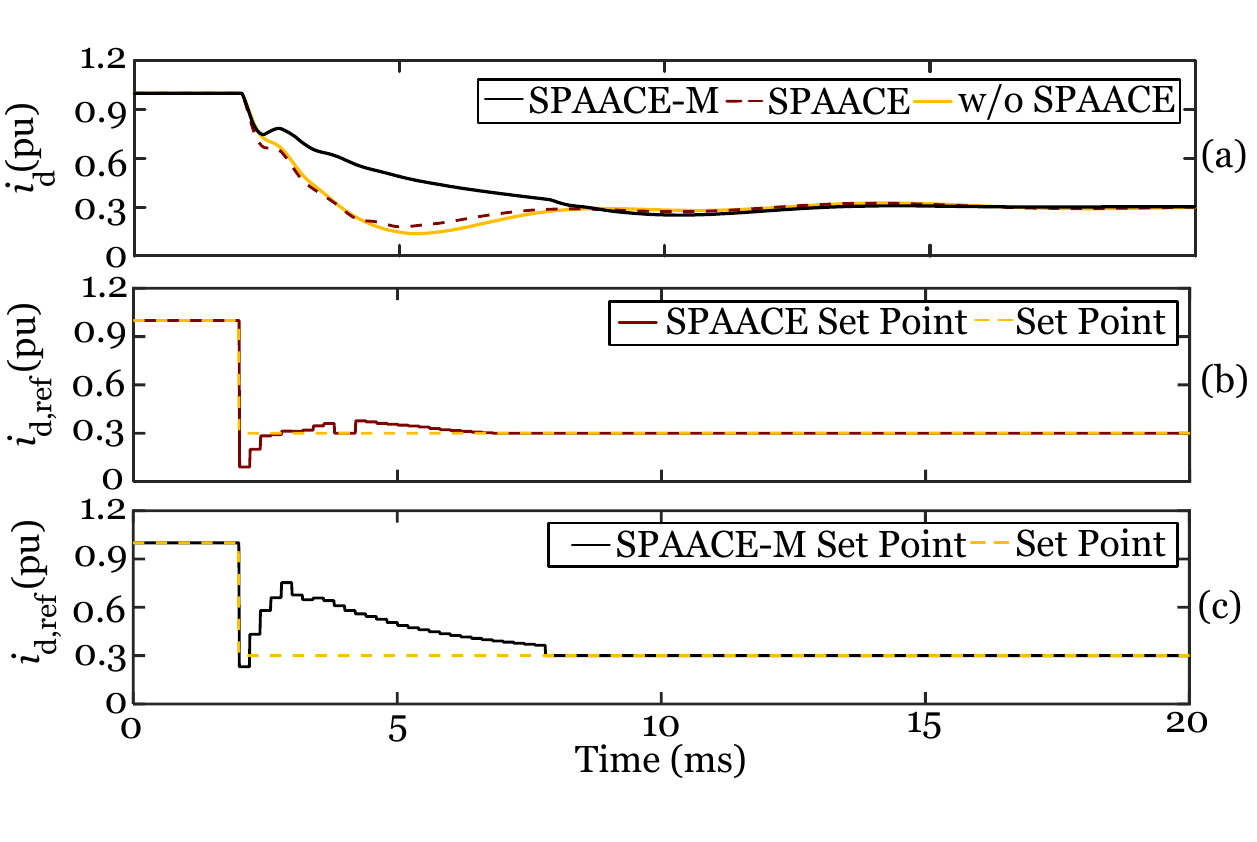}
\caption{Simulation results for Case~3.2: $i_d$ step change at $t=2~\rm{ms}$ from 1 to 0.3~pu with SCR = 1.}
\label{scr1}
\end{figure}

\subsection{Case 3: Impact of Grid SCR}
\subsubsection{Case 3.1: Set Point Decrease with Strong Grid}
Fig.~\ref{scr5} shows the simulation results for Case 3.1 tested on the power system shown in Fig.~\ref{single_bus_system}. The DER is operating in the steady state connected to the grid with an SCR of 5, which is a stiff grid. The sampling rate for SPAACE and SPAACE-M is 0.2~ms. At $t=2~\rm{ms}$, there is a step change in $i_{d,ref}$ from 1~pu to 0.3~pu. Fig.~\ref{scr5}(a) shows the $i_d$ response to the step change. In the base case, the current undershoots by 123.47\%. SPAACE is able to reduce the undershoot to 84.40\%, and SPAACE-M further reduces the undershoot to 19.22\%. Fig.~\ref{scr5}(b) shows how SPAACE modulates the set point to reduce the undershoot. Fig.~\ref{scr5}(c) shows the SPAACE-M modulated set point with a more gradual decrease to its final value while not increasing settling time.  

\subsubsection{Case 3.2: Set Point Decrease with Weak Grid}
Fig.~\ref{scr1} shows the simulation results for Case 3.2 tested of the power system shown in Fig.~\ref{single_bus_system}. The sampling rate for SPAACE and SPAACE-M is 0.2~ms. The DER is operating in the steady state connected to the grid with an SCR of 1, which is considered to be a weak grid. At $t=2~\rm{ms}$, there is a step change in $i_{d,ref}$ from 1~pu to 0.3~pu. Fig.~\ref{scr1}(a) shows the $i_d$ response to the step change. In the base case, the current undershoots by 54.47\%. SPAACE is able to reduce the undershoot to 40.40\%, and SPAACE-M further reduces the undershoot to 16.23\%. Fig.~\ref{scr1}(b) shows the SPAACE modulated set point, and Fig.~\ref{scr1}(c) shows the SPAACE-M modulated set point.

\subsection{Comparison Between Case Studies}

\begin{table}[!t]\footnotesize
\centering
\caption{  Comparison Results Table for Step Change Case Studies }
\label{table:comparison_table_stepchange}
\setlength{\tabcolsep}{1.6mm}\begin{tabular}{llllllll}
\toprule
\multirow{3}{*}{\parbox{ 1.6 cm}{\textbf{Case Study}}} & \multirow{3}{*}{\parbox{ 1.2 cm}{\textbf{Description}}}&
\multirow{3}{*}{\parbox{1cm}{\textbf{Algorithm}}}   &  \multirow{3}{*}{\parbox{1.25cm}{\textbf{Peak\\ Overshoot/\\Undershoot}}} & \multirow{3}{*}{\parbox{0.7cm}{\textbf{Settling Time (ms)}}} & \multirow{3}{*}{\parbox{0.8cm}{\textbf{Rise Time (ms)}}}
\\ 
\\
\\
\cmidrule{1-6}
 \multirow{6}{*}{\parbox{1.6cm}{Case~1: \\0.3~pu to \\0.7~pu set point increase }}& \multirow{3}{*}{\parbox{1.2cm}{0.2~ms\\ sampling rate}} & Base &    37.36\%	 &   14.59	  &  0.78 \\
    & & SPAACE &  24.27\% & 12.87 & 0.74 \\
  & & SPAACE-M &     10.42\%	 &   6.92	 &   2.34\\
   \cmidrule{2-6}
  & \multirow{3}{*}{\parbox{1.2cm}{3~ms\\ sampling rate}} & Base &    37.36\%	 &   15.6	  &  0.78 \\
   &  & SPAACE &  65.74\% & 27.78 & 0.74 \\
   && SPAACE-M &     18.45\%	 &   13.31	 &   0.93 \\
   \cmidrule{1-6}
 \multirow{6}{*}{\parbox{1.6cm}{ Case~3: \\1~pu to 0.3~pu set point \\decrease  }}& \multirow{3}{*}{\parbox{1.2cm}{Strong grid with SCR of 5}} & Base &    123.47\% &   16.87	  &  0.88 \\
    & & SPAACE &  84.40\% & 13.77 & 0.82 \\
   && SPAACE-M &     19.22\%	 &   9.57	 &   2.79 \\
   \cmidrule{2-6}
 &\multirow{3}{*}{\parbox{1.2cm}{Weak grid with SCR of 1}} & Base &    54.47\%	 &   13.49	  &  1.51 \\
  &   & SPAACE &  40.40\% & 13.07 & 1.47 \\
  & & SPAACE-M &     16.23\%	 &   10.28	 &   5.12 \\
\bottomrule
\end{tabular}
\end{table}

Table~\ref{table:comparison_table_stepchange} shows the peak value of overshoot and undershoot, settling time, and rise time of set point changes under the low- and high-sampling rate and under the weak and strong grid. 
SPAACE-M reduces the peak overshoot and peak undershoot in all scenarios compared to both SPAACE and the base case. 
Decreasing the sampling rate to 3~ms compared to 0.2~ms increases the overshoot for both SPAACE and SPAACE-M. As a result, using a higher sampling rate enhances the performance of SPAACE-M.
The results indicate that SPAACE cannot be used with a 3~ms sampling rate. 
Increasing the grid SCR from 1 to 5 increases the overshoot of SPAACE-M; however, grid SCR does not change SPAACE-M performance in terms of peak of overshoot and undershoot significanlty indicating that SPAACE-M can be used in both weak and strong grids.

Base case and SPAACE have almost the same rise time; however, SPAACE-M increases the rise time. Moreover, in slow sampling rate SPAACE-M shows similar rise time to the SPAACE and base case. 
SPAACE-M reduces the settling time in all scenarios compared to both SPAACE and the base case. 
SPAACE-M decreases the SPAACE settling time of an increasing step change by 46.23\% with a 0.2~ms sampling rate.
Decreasing the sampling rate to 3~ms instead of 0.2~ms increases the settling time for both SPAACE and SPAACE-M. 
Increasing the grid SCR from 1 to 5 decreases the settling time of SPAACE-M.
To conclude, SPAACE-M outperforms SPAACE in terms of overshoot, settling time, and the ability to use a lower sample rate with the trade-off of increasing the rise time.

\section{Conclusion}
A high sampling rate is needed for SPAACE to improve its performance, which is not always practical in the real world.
As the sampling frequency decreases, the performance of SPAACE degrades and shows overshoots larger than the base case.
This paper proposes the addition of a memory term to a SPAACE with a linear predictor. This method is designed to be implemented on existing DERs, improving their control without accessing any control settings. 
The proposed SPAACE-M algorithm reduces the overshoot, undershoot, and settling time to a step change as compared to SPAACE with the same sampling frequency, with the tradeoff of an increase in rise time. 
Moreover, SPAACE-M reduces the undershoot in three-phase shortcircuited faults compared to SPAACE.
 

\bibliographystyle{IEEEtran}
\bibliography{IEEEabrv,SP}

\end{document}